\documentstyle[12pt,aasms4]{article}

\journalid{}
\articleid{}

\begin{document}
\pagenumbering{arabic}

\title{Detection of CO emission $^{12}$CO J=$1\rightarrow0$ and
 $^{12}$CO J=$2\rightarrow1$ from the  Luminous
Blue Variable Star AG Carinae:\\ 
circumstellar envelope or disk?\altaffilmark{1}}

\author{A. Nota\altaffilmark{2,3}, 
A. Pasquali\altaffilmark{4},
A.P. Marston\altaffilmark{5}, 
H.J.G.L.M. Lamers\altaffilmark{6},
M. Clampin\altaffilmark{2} \\ \& 
R.E. Schulte-Ladbeck\altaffilmark{7}}

Accepted by {\it The Astronomical Journal}\\ 

\altaffiltext{1}{Based on observations collected at the European Southern
Observatory, La Silla, Chile.}
\altaffiltext{2}{Space Telescope Science Institute, 3700 San Martin Drive, 
Baltimore, MD 21218, USA; nota@stsci.edu, clampin@stsci.edu}
\altaffiltext{3}{On assignement from the Space Telescope Operations Division of
the European Space Agency}
\altaffiltext{4}{ST-ECF/ESO, Karl-Schwarzschild-Strasse 2, Garching bei
M\"unchen, D-85748, Germany; apasqual@eso.org}
\altaffiltext{5}{SIRTF Science Center, Mail Stop 314-6, Caltech,
Pasadena, CA 91125; tmarston@ipac.caltech.edu}
\altaffiltext{6}{Astronomical Institute of Utrecht University and
SRON Laboratory for Space Research, Princetonplein 5, NL-3584 CC, Utrecht,
The Netherlands; lamers@astro.uu.nl}
\altaffiltext{7}{Physics \& Astronomy Dept., University of Pittsburgh, 3941 
O'Hara Street, Pittsburgh, PA 15260, USA; rsl@phyast.pitt.edu}

\newpage
\begin{abstract}
We  present  the  first detection of $^{12}$CO J=2$\rightarrow$1 and $^{12}$CO 
J=1$\rightarrow$0   emission from the LBV AG Carinae.  We show that AG Carinae 
resides in a region which is very rich in molecular gas with complex motions. 
We
find evidence of a slow outflow of molecular gas, expanding at 
$\simeq$ 7 km s$^{-1}$. This emission appears  spatially unresolved. 
We argue that it is spatially
localised, rather than extended,  and possibly associated with the
immediate circumstellar region of AG Carinae. Both detected CO lines are
characterised by a pseudo-gaussian profile  of FWHM $\simeq$ 15 km s$^{-1}$,
indicating  a slowly expanding  region of molecular gas in
close proximity to the hot central star. We have explored two possible
scenarios to explain the observed profile: a circumstellar envelope, similar
to carbon stars, or a circumstellar disk.

The option of the circumstellar disk is preferable  because: 1) it is consistent
with additional independent indications for the existence of wind asymmetries
in close proximity to the  central star,  found from spectropolarimetry and
analysis of the UV and optical line profiles, and 2) it provides  the
conditions of density and shielding necessary for the survival of the CO
molecules in  proximity to such a hot star (T$_{eff}$ $\simeq$ 14000 K - 20000
K). In the assumption that the CO emission originated when AG Carinae was
in an evolved state, we derive a lower limit to the CO-mass of 6.5
$\times$ 10$^{-3}$ M$_{\odot}$. We also estimate that the CO fraction is 
$\simeq$ 2.3 $\times$ 
10$^{-3}$ of the total mass of molecular gas, which then would amount to 2.8 
M$_{\odot}$.
This is smaller, but still comparable with the mass of ionized gas present in 
the circumstellar
environment (4.2 M$_{\odot}$), with the implication that the molecular gas 
fraction can contribute significantly to the overall  mass lost from the 
central star in its post main sequence evolution.

\end{abstract}

\section{Introduction}

Massive stars spend their H-burning phase as O stars and become,  after a  few
million  years, Wolf-Rayet (WR) stars undergoing core He-burning. Mass loss  
plays a key role in driving the stellar evolution from the main sequence to the
WR  stage, especially during  the intermediate phase of Luminous Blue Variable
(LBV).  Luminous Blue Variables  have been recognized to be at the point of
core H-burning  exhaustion (Pasquali et al. 1997). In the H-R Diagram (HRD)
they are located  close to the Humphreys-Davidson  (HD) boundary limit
(Humphreys \& Davidson 1979),  which represents an empirical instability limit
to the evolution of stars more  massive than 40 M$_{\odot}$. They  display
large and irregular  spectro-photometric variability by transiting between hot
visual minima phases (T$_{eff}$ $\sim$  20000 - 30000K) to cooler visual maxima
phases (T$_{eff}$ $<$ 10000K). Their  variability is also characterized by 
violent eruptions, with visual brightness increases  of 3 magnitudes or more,
and extreme mass loss (up to several solar masses of  material ejected in the
surrounding medium).  In between such dramatic outbursts, LBVs  still lose mass
at  high rates -- typically 10$^{-5}$ - 10$^{-4}$ M$_{\odot}$/yr (e.g. 
Leitherer
et al. 1994).

The relics of these eruptions are spectacular circumstellar nebulae, such as
the one  observed around Eta Carinae. A systematic investigation of
LBVs in the Galaxy and in the Large Magellanic Cloud (Nota  et al. 1995) has
shown that the majority of the known LBVs is surrounded by an ejected nebula,
whose kinematical and chemical properties may be used to constrain the mass
loss history of the central star. Their morphology  suggests they are
shaped by wind interactions. The nebular dynamics indicates that the ejection
timescale is comparable to the evolutionary lifetime of the LBV phase. Their
chemical composition has been used to infer the evolutionary status of the
central star at the moment of the ejection (Lamers et al. 2001). However, these evolutionary
diagnostics are mostly derived from optical data. So far, LBVs have been little
studied at infrared  or submillimeter wavelengths  although  these  spectral
regions  provide a number of diagnostics probes of the nebular physical
conditions which are not available at optical or UV wavelengths. A great wealth
of information  can be derived from  such observations  on the presence of dust
or molecular gas in the close surrounding  of these hot stars, to complement 
the optical data and to help determine
the total amount of mass lost during the LBV phase. This piece of information,
presently incomplete, is crucial to constrain the theoretical models which
currently fail to reproduce the transition between O and WR stars (Langer et 
al.
1994). 

In the past few years, near-IR imaging and spectroscopy both from the ground
and more recently with ISO (Waters et al. 1998a,b; Voors et al. 1997; Morris et
al. 2000) have greatly expanded our understanding of the properties, formation
and survival mechanisms of dust in  hot stellar environments (T$_{eff}$
$\simeq$ 20000 - 40000 K).  For the first time  it has been
possible to study the spatial distribution of the dust, to quantify the nebular
dust mass with some accuracy, and to investigate the dust - gas correlation. 
In addition,  the spectroscopic investigation of the dust features
has contributed to better characterise the dust properties and  has provided
additional independent evidence that these stars were cool and extended 
(mimicing a Red Supergiant - RSG) during the ejection of the nebula (Lamers et al. 2001).

Presence of molecular gas around  hot, evolved stars had  unexpectedly been 
found  in
1988 by McGregor et al. in a survey of blue supergiants in the LMC. McGregor et
al. unambiguously detected  the first overtone band heads of CO at 2.3 $\mu$ in
six objects. Given the very low CO dissociation energy (11.1 eV), the presence
of CO emission in close proximity to O and B supergiants characterised by fast
stellar wind (V$_{inf}$ $\simeq$ 200-400 km s$^{-1}$) and strong radiation 
fields has been difficult to explain. One possible explanation is that the
CO emission is spatially extended, and far from the star. If the star has
already experienced a RSG phase,  the CO emission could be a tracer of the
previous cool and dense wind which has been compressed and excited by the
interaction with the present hotter stellar wind. Alternatively, the CO
emission may be  localised  and close to the star. McGregor et al. (1988) had
tentatively identified the CO emitting region with  a circumstellar disk which
is able to shield the CO from the stellar radiation and ionized wind. 
Localised emission can also be explained, as  is observed in the case of 
cooler
giants and supergiants (such as S and C
stars), by the survival of the  CO  in  an outer layer of the stellar envelope 
and its excitation by the stellar radiation field (Knapp \& Morris 1985; 
Olofsson
et al. 1988; Heske 1990). This explanation is only acceptable if the star mimiced a
RSG during the ejection of the nebula. Unfortunately, the observations by McGregor
did not  have the spatial resolution necessary to discriminate among the possible
scenarios. 

Among the luminous supergiants which McGregor et al. (1988) observed, there was
the galactic LBV AG Carinae. AG Carinae is one of the most luminous LBVs, which
is located in the  HRD   very close to the HD
limit. Its distance is $6 \pm 1$ kpc
and its luminosity is $1.7~10^6$ L$_{\odot}$ (Humpreys et al. 1989).
The effective temperature is variable between about 9000 K and
25000 K, and so the radius changes from 70 to 500 R$_{\odot}$
(Lamers 1986; Voors et al. 2000). It is surrounded by a very bright, extended nebula which has
been known since the 1950's (Thackeray 1950). The nebula is composed of ionized
gas and dust. When imaged in the light of H$\alpha$, it exhibits an elliptical
shell-like morphology (30$''$ $\times$ 40$''$ in size). The nebular lines
indicate that the nebula is expanding at an average velocity of 70 km s$^{-1}$
(Smith  1991). In the light of the line free continuum, which displays the stellar
radiation scattered by dust, 
the AG Carinae nebula appears quite different (Paresce \& Nota 1989):  
the dust grains are
distributed in a jet-like feature which  HST has  resolved into a
myriad of clumps and filaments with an overall bipolar structure (Nota et al.
1996). Recent ISO observations indicate that the dust grains are mostly
crystalline olivine with an emission peak at 33.8 $\mu$m (Waters et al. 1998).
  McGregor et al.
(1988) did not detect any CO overtone emission in AG Carinae.  They found
little or no emission from hot dust in the near-infrared, but surprisingly
detected strong far-infrared emission from cool dust, and  inferred the action
of a strong stellar wind, which possibly removed  the dust from the inner
region. We decided to revisit the question of the presence of molecular gas in
the circumstellar environment of AG Carinae, for a number of reasons:   AG
Carinae is the ideal target for a neutral material study, given its proximity
and  the large spatial extension of the nebula. The nebula is rich in  gas and
dust and displays a strong far-infrared eccess. We have therefore decided to
start an investigation aimed at detecting and characterising the properties of 
the
neutral material around AG Carinae, with the  objective to: 1) detect
CO emission from molecular gas, 2)   understand whether the emission is 
localised or spatially extended, and 3) determine the amount of molecular gas
in the circumstellar environment.  We have carried out this program  with the
SEST telescope in the submillimeter range, studying the transitions  $^{12}$CO
J=$2\rightarrow1$ and $^{12}$CO J=$1\rightarrow0$  and obtaining CO emission
maps of the entire nebula. The observation strategy and data reduction
procedures are described in Sect. 2, and the results of our investigation are
presented in Sect. 3. Discussion and conclusions follow in Sect. 4 and 5,
respectively.

\section{Observations and Data Reduction}

Observations of the region around AG Carinae were made with the Swedish-ESO
Submillimeter Telescope (SEST) in February 1997, March 1998 and February 1999.
In February 1997 observations were made with  the 3 mm receiver and High
Resolution Spectrometer (HRS) covering the $^{12}$CO J=$1\rightarrow0$
(115.271GHz) emission line. This provided 2000 frequency channels covering an
86MHz bandwidth. The frequency resolution of this system is 80kHz, providing a
velocity resolution of approximately 0.22 km s$^{-1}$. At this wavelength SEST
has a beamwidth of approximately 45$^{\prime\prime}$. Single beam spectra were
taken along an east-west strip from 12$^{\prime}$ E of AG Carinae to
15$^{\prime}$ W of AG Carinae using 45$^{\prime\prime}$ steps.  A total of 36
spectra were acquired, with typical on-source integration times of 30 seconds.
A dual beam switching method was used in which the source was alternately
placed in the two beams to eliminate baseline ripples. The beam throw was
11.6$^{\prime}$.  For several of the positions, emission in the off-source
beams affected our final profile. The corresponding spectra were therefore 
discarded during our analysis. 

Additional observations were taken with the same configuration on February 28,
1999, with the objective of extending the spatial coverage to  30'  to the East and
to the West of AG Carinae. Typical exposure  times were 30s to 2m.  A smaller map
in declination was performed in the same configuration, from 0 to 10$'$ North and
South, with pointings at  0, -2$'$, -10$'$, +2$'$, +10$'$. In Figure~1
we show a large field image centered on AG Carinae, generated from  POSS plates, on
which we have drawn the  region covered by  the $^{12}$CO J=$1\rightarrow0$
observations.

Observations were made of the southwest quadrant of AG Carinae nebula in March
1998 using the 230GHz receiver with HRS centered on the $^{12}$CO
J=$2\rightarrow1$ (230.537GHz) emission line. We concentrated on this spatial
region first, because optical images show the highest concentration of gas
emitting and dust scattering stellar light. The complete mapping of the
nebula was then  achieved in  February 1999, with the other three quadrants
covered. At these frequencies, our velocity resolution is 0.11 km s$^{-1}$ and
the effective beam width of the telescope is 23$''$.  The
observations  of the SW quadrant covered a regular grid of 36$''$
$\times$ 36$''$ at 12$''$ intervals, starting at the
position of the star. The SE quadrant covered a grid of nine observations, to
24$''$ S,  also spaced by  12$''$. The NE quadrant covered nine pointings
extending to 36$''$ E, symmetrical to the SW quadrant, and finally the NW
quadrant covered nine pointings, also spaced by 12$''$.  In
Figure~2 we show all pointings overlaid on  a  ground based
coronographic   image of the AG Carinae nebula, taken in the light 
of H$\alpha$ (Nota
et al. 1992). Several on-source observations were made at each grid position
with 30 seconds integration each.  Summation of these spectra provided total
on-source integration times  5 to 10 minutes at each of the grid positions. 
Elimination of baseline ripples was achieved by position switching to an area
of the sky apparently devoid of CO emission-lines.

For both sets of observations, pointing to better than 3$''$ was
achieved through observations of bright SiO masers, e.g. VY CMa.

Intensity calibration, resulting in an antenna temperature T$_A^*$, was
provided by the chopper wheel method and a set of internal
calibrators.  This was done on-line as the data was collected at the
telescope.  Linear drifts in frequency were also removed during each
calibration through the use of a frequency comb (a set of narrow lines
with well defined frequencies).  To convert the intensities to main
beam brightness temperatures (T$_{mb}^*$), the antenna temperatures were
divided by the main beam efficiency of the telescope (0.7 and
0.5 at 115 and 230GHz respectively).  For both sets of observations
constant and/or residual sinusoidal baselines were removed, as
necessary, from each of the final CO profiles. The resulting profiles
were then plotted on a velocity scale, after being reduced to the local
standard of rest (V$_{lsr}$).

\section{CO maps and CO-line profiles}

The $^{12}$CO J=$2\rightarrow1$  and the  $^{12}$CO J=$1\rightarrow0$ profiles
taken at the star position are plotted in Figure~3. They are
both characterised by a broad component, with a FWHM of $\simeq$ 15 km s$^{-1}$
centered on the star, and by two narrow components, which in the $^{12}$CO
J=$1\rightarrow0$ profile have a FWHM of 2 km s$^{-1}$ and 1.2 km s$^{-1}$
respectively, and are separated in velocity by approximately 10 km s$^{-1}$.  
\\
The first goal of this work is to establish which components are intrinsic to 
AG
Carinae and which are generated by the underlying interstellar region. 

\subsection{The $^{12}$CO J=$1\rightarrow0$ map}

We investigated the circumstellar region around AG Carinae, by
performing a  map  in right ascension of the $^{12}$CO J=$1\rightarrow0$ line,
from  30$'$ West  to 30$'$ East, and a smaller map in declination, from 0 to
10$'$ to the North and to the South. The  map is illustrated in
Figure~4, where selected pointings in the  North, East
South and West regions are shown, and compared with the observation on the star.
 The sampling of the observations is
different in different regions: every 3$'$ in the East, every 2$'$  in the West, and selected
pointings  in the North-South direction (0, -2$'$, -10$'$, +2$'$, +10$'$). Some
of the pointings along the East-West direction  show the two {\it "narrow"} components  seen at the star position,
although at different levels of intensity,
broadening  and separation.  They are easily visible in the East region, although
with remarkable variations in relative intensity (eg. pointings at +12$'$ and +20$'$
in Figure~4). They become very faint in the
West region, and almost disappear beyond detection westward than 14$'$.  They
are visible  in the South, but they are very weak in the North. In 
Figure~5, we have illustrated the motion of these two components in the
East - West direction. Their velocity is reported, referenced to the LSR, as a
function of position  with respect to the star (East to the left and West to
the right). The overall motion is consistent with  parallel sheets of neutral
material at constant radial
velocity and very different emissivity from point to point. From the line
profiles in the maps, it
is also clear that other narrow components are present, indicating that {\it the source
of this narrow  CO component
 is in the complex motions of the underlying molecular
gas, and is  not physically associated with AG Carinae}.

The {\it "broad"} component appears to have a different  . It is
strongest in  the pointing centered on the star and somehow visible, although
at lower intensity levels, in the spectra at $+$4$'$ and $+$2$'$. However, this
{\it "broad"} component disappears completely beyond a few arcminutes from the
central star, as it can be seen in Figure~4.  We have fitted a 
 gaussian of FWHM $\simeq$ 12  km s$^{-1}$ to  the
broad component in each of the pointings in which it is visible, other than the
central star, to estimate its intensity in the pointings
adajacent to the star. We find that the {\it  broad } component  observed on the star
is approximately a factor two brighter than the average of the adjacent
pointings in which the {\it broad} component, is at some level, visible
(Figure~6). 

It is important to understand whether the emission  is associated with the AG
Carinae system. From the observations presented so far  it is clear that the
emission is maximum in correspondence to AG Carinae, and that it extends over 
a region which is possibly a few arcminutes in size.  However, it  is not
possible, on the basis of these observations alone, to conclude whether the
emission originates from the star and close circumstellar environment or from
the surrounding nebula, which is rich in dust, and most plausibly could also
contain molecular gas. Unfortunately the spatial resolution of  the $^{12}$CO
J=1$\rightarrow$0 observation is $\simeq$ 45$''$, which is roughly
corresponding to the spatial extent of the  circumstellar nebula. The
combination of the spatial resolution (45$''$) with the spatial sampling (2$'$)
does not allow us to assess, from these observations alone, the origin of the
detected  emission. 

We  then performed a second map in the $^{12}$CO J=2$\rightarrow$1 line, with  the
expectation to be able to better define the nature of this broad emission. In fact,
in this  observing mode the spatial resolution  of SEST is almost a factor two
better (beam width = 23$''$, compared with 45$''$).

\subsection{The $^{12}$CO J=$2\rightarrow1$ map}

The central section of the grid of $^{12}$CO J=2$\rightarrow$1 spectra covering the  AG
Carinae   nebula  is shown in Figure~7. The SW quadrant of the nebula
is of particular interest because it presents  the highest concentration of gas 
and dust. This is the  region where we would expect to detect  the presence of
molecular gas and  therefore, CO emission. Spectra of this region have been taken
in 1998  with a spacing of 12$''$ in both RA and  DEC. Additional maps of the
remaining three quadrants were obtained a year later with the same observing
configuration.  The total spatial  coverage of the $^{12}$CO J=$2\rightarrow1$ 
map  is $\simeq$ 72$''$ in the two  directions (E-W, N-S). The spectra cover the
velocity range from -2 km~s$^{-1}$ to 51 km~s$^{-1}$ and a beam temperature
T$^*_{mb}$ range from -0.4 K to 2.3 K. All spectra acquired at each position have
been summed. As  can be seen from the map, the $^{12}$CO J=2$\rightarrow$1 profiles
are characterised by a complex  structure of sub-peaks where a number of recurrent
features can be identified.  For example, the two dominant narrow components
observed in the $^{12}$CO J=1$\rightarrow$0 spectra are detected in all $^{12}$CO
J=2$\rightarrow$1 profiles. Their morphology seems to change as function of
position. In the South, the blueshifted component develops a broad and complex profile, with a
number of  sub-peaks.  The redshifted component increases  in intensity with
distance from the star, reaching a peak at $\Delta$ DEC = -24$''$.  Additional, fainter,
emission sub-peaks  are  found  between the two  dominant components, and we notice
they tend to merge with the two  dominant narrow components at increasing distance
from the star. 

  Throughout the map, the  same observational features are found, namely:  1) the 
two recurrent narrow emissions which eventually blend with secondary peaks, 2) a
broad component well visible in the pointing  on the star and  in many 
adjacent positions.

In order to better characterise the evolution of  the emission components as a
function of spatial position, we have applied a multi-gaussian fit to all the
spectra in the  $^{12}$CO J=2$\rightarrow$1 SW map, to: 1)
identify the sub-peaks detected at some of the offset positions, and 2) 
track their motion as a function of position with respect to the star.  The
velocity-position map is shown in Figure~8, where we have
identified 10 separate components in the $^{12}$CO J=2$\rightarrow$1 profile
seen on the star (top panel)  and tracked them at increasing distances from
the star, in the SW quadrant.  Their velocities, reduced to the LSR, are
plotted  as a function of distance from the star in the bottom panel of
Figure~8. The filled and  open circles indicate the two  dominant
narrow  components - already discussed  - at V$_{lsr}$ = 18.5 km s$^{-1}$ and
29.7 km s$^{-1}$, respectively. Five additional components are present in {\it
all}  positions observed: these are at V$_{lsr}$ = 11 km s$^{-1}$, 18.5
km s$^{-1}$, 21.5 km s$^{-1}$, 29.7 km s$^{-1}$ and 41 km s$^{-1}$. Other
components  are detected, which have a  more limited spatial extent: a feature 
at 
V$_{lsr}$ = 13.5 km s$^{-1}$ appears only at  distances  larger
than 17$''$ from the star while a component at V$_{lsr}$ =  26 km s$^{-1}$ 
disappears at
distances larger than 38$''$. Two other components, at V$_{lsr}$ = 27 and 28
km s$^{-1}$, develop at  distances of 24$''$ and 27$''$,  respectively. 
Finally,
a component at V$_{lsr}$ = 33.5 km s$^{-1}$ is not detected at distances larger
than 12$''$. It is interesting to notice that  {\it all detected components}
have constant  velocity over the entire spatial range. 
This shows that the CO producing the narrow components is most likely in 
sheets of neutral material moving at constant velocity in the
background/foreground of the AG Carinae system. 

The most  intriguing feature remains, however,  the broad component observed in
the spectra taken on and around the central star. It is still detected, but
much less prominent,  in the spectra on regions further away from the star
 (eg. (+12$''$, -24$''$), and (+12$''$,
-12$''$) most likely due to the fact that part of the
beam (FWHM = 23$''$) still subtends the inner region. It seems unlikely that the emission
is originated by the AG Carinae optical nebula.  In fact, the AG Carinae 
nebula is $\simeq$ 35$''$ in diameter. The brightest regions are located
$\simeq$ 12$''$ from the star. If the CO emission were originating from the
gaseous nebula, we would  expect to detect little CO emission in the spectrum
obtained on  the star, and strong emission at  distances 10-20$''$ from the
star. We observe exactly the opposite: the emission is stronger and best
defined in the spectrum taken on the star, and is much less prominent as soon
as we depart from that position. This indicates clearly that  the  source of
the  broad CO component is in the close circumstellar environment of AG Carinae.

An independent confirmation of  our conclusion comes from
kinematical considerations: we have fitted  a multi-gaussian function to the
$^{12}$CO J=2$\rightarrow$1 profile acquired on the star (0, 0), to disentangle
the broad component from the two narrow components which are detected at that
position. The results of the fit are shown in Figure~3 (bottom 
panel).
We have measured  the broad component  to have a FWHM  of $\simeq$ 15 km
s$^{-1}$, a peak velocity of 26.5 km s$^{-1}$, and a 
peak T$^*_{mb}$ of 0.8 K.   For comparison, in Figure~3 
(upper panel) we show the $^{12}$CO
J=1$\rightarrow$0 line profile, 
which is very similar in morphology. In the case of the $^{12}$CO
J=1$\rightarrow$0 line we have derived a peak V$_{lsr}$ of 24.5
km~s$^{-1}$, a peak T$^*_{A}$ of 1.5 K and a FWHM value of 14 km~s$^{-1}$. This
resulting expansion velocity of the outflow ($\simeq$ 7 km s$^{-1}$) is much lower than
the expansion velocity of the optical nebula  ($\simeq$ 70 km s$^{-1}$, Smith
et al. 1991). This difference in velocity again indicates that the CO emission
does not originate at the location of the optical nebula, at a distance of $\simeq$20$''$
from the central star. The location and the velocity of the CO broad component show that
 the CO emission comes from a region close to the star, with a small outflow
velocity of $\simeq$ 7 km s$^{-1}$. Of course this broad component could be due to a
foreground or background cloud that happens to coincide with the position of the
star. This is very unlikely, however, for the following reasons: first, because of
the low probability of such a coincidence, and second,  because of the larger 
width
of the CO broad line compared with the other  narrower interstellar CO components.

\section{Discussion} 
Observations of the first overtone band of CO at 2.3 $\mu$m  in AG Carinae were
attempted by McGregor et al. (1988).  No CO emission was detected in the
infrared spectrum of AG Carinae or in Eta Carinae and P Cygni, the
three most well known galactic LBVs. Our detection of a broad emission in the
submillimeter $^{12}$CO J=2$\rightarrow$1 and $^{12}$CO J=1$\rightarrow$0 lines
is, therefore, the first to date.  This type of emission is typically found in 
carbon stars, which lose large amounts of matter in the form of a slow stellar
wind, and typically form circumstellar envelopes of gas and dust. The physical
conditions of temperature and density in these envelopes are such  that 
molecular species can  form and survive, and be shielded from the dissociating
UV radiation field (Oloffson 1988, Oloffson et al. 1993a,b)

Carbon stars come from low-mass progenitors and  follow a  very different
evolutionary path  compared with the  much more massive counterparts discussed
here. However, it is interesting to notice that an inspection of the CO line
profiles collected by Olofsson et al. (1993) from  their large sample of
galactic bright carbon stars shows that the broad component observed in AG
Carinae quite closely resembles similar features observed in the spectra of
several stars in their sample.  In general, the observed line profiles were 
found to lie  in the middle of two extremes: the rectangular profile 
and the parabolic profile.  

In determining the amount of CO, we have to make an assumption
about its spatial and velocity distribution. 
The broad emission observed in AG Carinae closely mimics the standard 
parabolic
profile (cf. Figure 3). Such a parabolic profile can be due  to an optically thin outflow, if the beam size is smaller than the
angular extension of the source.
This option is excluded on the basis that we believe the emission to be  unresolved.
The parabolic profile can also be due to:
\begin{itemize}

\item {\it a)} an optically thick spherically symmetric outflow at constant velocity,
\item {\it b)} an optically thin wind with a non-spherical outflow or a velocity that is
not constant with distance.
\end{itemize}
 The presence of CO in close 
proximity to these
very hot stars implies that the neutral material must be shielded from the 
stellar radiation.
This is possible if
the outflow is mainly in a disk (option b). This disk could be either
 gravitationally
bound,
i.e. a Keplerian disk, or an outflowing disk. As shown by Bjorkman (1998)
and Kraus \& Lamers (2002)
effective shielding occurs in  disks around B[e]-stars, so that
H remains neutral very close to the stars and CO can form at a distance
of only a few stellar radii. 

Moreover, the  study of the chemical abundances in  LBV nebulae 
(Lamers et al. 2001) and  the currently accepted scenarios for the 
nebula ejection (Langer 1997)   indicate that rotation must play an
important role in the mixing and the ejection of the nebula.
Therefore, we cannot use the optically thick spherical
approximation to estimate the amount of CO mass (option a). Instead
we will use the {\it  optically  thin approximation} to estimate a {\it
lower limit} of the mass.
The line luminosities of the 2.6mm and the 1.3mm lines are derived
from the source monochromatic brightness $B_{\nu}(T)$ per steradian,
 where the brightness temperature T 
(= T$_{mb}^*$) is the antenna temperature
T$_A^*$ corrected for the telescope efficiency. The net line flux is computed
from the `on the star' broad component after subtraction of the 
mean interstellar contribution. The monochromatic brightness is actually the integral of 
the residual temperature profile as a function of frequency for the CO 
$J=1\rightarrow0$
line and the beam size of the telescope:
\begin{equation}
F_{2.6mm} = Beam \times \Sigma B_{\nu}(T)\Delta\nu
\end{equation}
Therefore, the luminosity of the $J=1\rightarrow0$ line turns out to be:
\begin{equation}
L_{2.6mm} = 4\pi D^2F_{2.6mm} = 1.3 \times 10^{-3} L_{\odot}
\end{equation}
Similarly, we find for the 1.3mm line,
\begin{equation}
L_{1.3mm} = 2.3 \times 10^{-3} L_{\odot}
\end{equation}

If the emitting region is optically thin, the line luminosity can be
expressed in terms of the total number of emitting CO molecules 
in the  $J=1$ level as $L_{2.6mm}=N_{J=1} A_{J=1,0} h \nu$  and
similarly for the $J=2\rightarrow1$ line,  where
the Einstein coefficients $A_{J=1,0}=
6\times 10^{-8}$ s$^{-1}$ and $A_{J=2,1}=
7\times 10^{-7}$ s$^{-1}$. The population of the $J=1$ level 
can be expressed in terms of an excitation temperature as
\begin{equation}
N_{J=1} = N_{J=0} \frac{g_u}{g_l} exp(-\chi/kT_{\rm exc})
\end{equation}
with $g_u=3$ and $g_l=1$ and $\chi = 4.77 \times 10^{-4}$ eV.
The excitation temperature is at least as high as that of the 2.7 K
background radiation. This implies that N$_{J=0}$ $<$ 2.6 
N$_{J=1}$ and N$_{tot}$ $<$ 3.6 N$_{J=1}$.  We can estimate the
excitation temperature from a comparison of the line luminosities
 of the CO$_ {J=2\rightarrow1}$ and the CO$_{J=1\rightarrow0}$
 lines, which is  L$_{2\rightarrow1}$/L$_{1\rightarrow0}$ = 1.7.
 With the parameters given above, we find that N$_{J=2}$/N$_{J=1}$ =
 0.090
 which corresponds to an excitation temperature between these two
 levels of 3.8 K. Note that this is the excitation temperature
 produced by collisions with H$_2$ molecules, and not the gas
 temperature which can be much higher.
  Adopting this value for both excitation steps we
 find that N$_{tot}$ = N$_{J=0}$ +  N$_{J=1}$ + N$_{J=2}$ = 2.52
 $\times$ N$_{J=1}$. We will adopt this ratio for the derivation of
 the lower limit to the CO mass.

 From the values of L$_{2.6 mm}$ and L$_{1.3 mm}$ we find that
 the total number of CO molecules required to produce these line
 strenghts is 2.3 $\times$ 10$^{53}$. This implies a minimum CO mass
 of 5.2 $\times$ 10$^{-3}$ M$_{\odot}$ ==> 5.4 $\times$ 10$^{-3}$ M$_{\odot}$.

We can estimate the total amount of molecular gas by quantifying the
present fraction of CO. Studies of the CNO abundances in the nebulae of LBVs 
by Smith et al. (1998) and  Lamers et al. (2001) show that the nebula of AG Car
consists roughly of 80\%  processed material in which C is
severely depleted, and 20\% of unprocessed material. This suggests
that the mass fraction of  C is depleted by about a factor 5 relative to the
initial composition, for which the mass fraction of C is 4.9 $\times$ 10$^{-3}$
in Population I gas.
Oxygen is also depleted, but the C/O ratio will still be
smaller than unity so that the CO abundance is set by the C abundance. If all
the C is locked in CO, which is a reasonable assumption for a shielded region,
the mass fraction of $^{12}$C$^{16}$O is  28/12 times the mass fraction of C.
In the assumption that the CO emission was generated when AG Carinae was an  evolved
object, the CO mass fraction would then be of  $(1/5) \times (28/12)\times 
4.9\times 10^{-3}=2.3 \times 10^{-3}$. Accounting for this fraction we find a
total  mass of the molecular region of 2.8 M$_{\odot}$. With this estimated minumum
mass, and a maximum radius of the unresolved CO-emitting region of about 30 $''$ in
diameter, we find that the optical depth of the CO lines is $\tau_{J=1\rightarrow0}
\simeq 8$ and $\tau_{J=2\rightarrow1} \simeq 4$. So the lines are marginally optically
thick and the mass estimate is indeed a lower limit. The H$_2$ density,
derived from the CO column density and the ratio n(H$_2$)/n(CO)$\simeq 6
\times 10^3$, is at least $5 \times 10^2$ mol cm$^{-3}$.

This cool gas is most likely concentrated in 
the outer regions of a 
circumstellar disk. 

Such a disk might be the result of a strong equatorial mass loss,
possibly combined with wind compresssion, similar to the disks  of
B[e] supergiants (Zickgraf et al. 1986; Bjorkman 1998; Cassinelli 1998)
 or due to an
outburst when the star has reached its $\Omega$ limit, which is the
rotationally modified Eddington limit (Langer 1997).
In the case of AG Carinae, 
some circumstantial evidence already exists that 
the star
might host a gaseous disk (which is required to shield the CO in its outer layers
from the stellar radiation), based on several independent observational factors:
\begin{itemize}
\item  the bipolarity of the circumstellar nebula; especially in the light of 
the
continuum, the ejected nebula surrounding AG Carinae displays a remarkable
axisymmetry (Nota et al. 1996) which could have been produced by a strong 
stellar wind
interacting with a pre-existing equatorial enhancement;
\item the variability in the intrinsic polarization ($\simeq$ 1.2$\%$).  Very 
large
variations  in AG Carinae's linear polarization   have been observed with time 
at UV
and optical wavelengths (Schulte-Ladbeck et al. 1994; Leitherer et al. 1994).
The polarization is found to vary along a preferred position angle  which is
co-aligned with the major axis of the circumstellar nebula. The authors 
interpret
this finding as the presence of a strong anisotropy in the stellar wind.  
\item the coexistence of
different wind components in the UV line profiles. The UV spectra  show  both 
presence of recombination lines   due to a slow, dense wind and  UV P Cygni 
profiles indicative of a faster, less dense wind (Leitherer et al. 1994). This 
two
component
wind structure is very reminiscent of the wind conditions in B[e] stars where 
the
slow, dense wind is usually referred to as a circumstellar disk (Zickgraf et 
al. 1986; Kraus \& Kr\"{u}gel 2002). 
\end{itemize} 
The comparison with the B[e] supergiants is also interesting. B[e] stars share 
approximately the same location in the HRD with LBVs, although at slightly 
cooler
temperatures. B[e] supergiants show clear evidence for a fast low density wind
with V $\simeq$ 1000 km s$^{-1}$, that can be seen in the UV resonance lines,
and a more slowly expanding equatorial disk with V $\simeq$ 100 - 200 km 
s$^{-1}$
that can be seen in the Balmer lines, in the low excitation permitted emission
lines of the metals, and in the forbidden emission lines of [FeII] and [OI].  

Zickgraf et al. (1986) have explained the observed bimodality in terms
of a two-component wind, developing along the polar axis and on the
equatorial plane of the stars. Such a structure can be explained by the 
rotation induced
bi-stability model (Lamers \& Pauldrach 1991; Pelupessy et al. 2000). 
The polar wind is a
radiation-driven wind, hot but not dense, expanding at higher speed.
It is responsible for the strong P Cygni emissions of the UV resonance lines. 
The equatorial outflow, which is also radiation-driven, but by ions of lower 
ionization stages
than the polar wind, is cooler and denser, moving at
lower velocity. The conditions in the dense equatorial outflow allow shielding 
of the heavy
elements and are met by an  equatorial enhancement  of the stellar wind,
which is also referred to as a disk. This disk is possibly further compressed 
by
the wind compressed disk mechanism (Bjorkman \& Cassinelli 1993; Lamers \&
Cassinelli 1999, Chapter 11). The circumstellar disk
prevents dust from being destroyed by the stellar radiation and  may
well explain the  characteristic strong infrared excess observed in B[e] 
supergiants due to thermal emission from dust at T $\simeq$ 1000 K
(Bjorkman 1998).

AG Carinae is not the only LBV for which the possible presence of a disk has
been indicated by a number of independent observations. HR Carinae for 
example, which is
to date the most plausible LBV candidate  for having a  disk, shares some of 
the 
same observational evidence: 1) a bipolar nebula, 2) CO overtone emission 
detected
in  the  infrared spectrum (McGregor et al. 1988),  indicative of a  cool
(a few thousand K) and dense ($\simeq$ 10$^{10}$ cm$^{-3}$) region of 
circumstellar
gas, 3)  evidence for presence of intrinsic
polarization, from the change of polarization at the H$\alpha$ emission
line with respect to the continuum (Clampin  et al. 1995). These results 
imply that the  scattering 
material  in close proximity to the star is also not distributed with a 
spherical
symmetry. In addition, in similarity with AG Carinae,  the intrinsic 
polarization
vector turns out to be  coarsely  perpendicular to the nebular major axis.

It is interesting to understand why McGregor et al. (1988) failed to detect
any near-IR CO overtone emission in AG Carinae, while detecting it in HR 
Carinae.
McGregor et al. observed HR Carinae  four  times between 1984 and 1986. In 
March
and June 1984 they detected the three CO overtone bands at $\lambda$ 2.295,
2.324, and 2.354. In March 1985 and February 1986 they reobserved the star, 
but could 
only obtain an upper limit to the flux. Morris et al. (1997) reobserved
HR Carinae in November 1995  with ISO and again detected presence of the three
CO band heads.  They found that a combination of a bi-stable wind
with rapid rotation or wind compression forming a disk would meet the density 
and
shielding required to explain the observed CO flux. In addition, they found 
that
the variability of the CO overtone emission was in phase with the variations
in the V magnitude,  that is  CO overtone  would be detected when the light 
curve was at a minimum.  Morris et al. (1997) explained this observed 
variability with optical depth variations as predicted by the theory
of bi-stable line-driven winds (Pauldrach \& Puls 1990), and for bistable disks by
Lamers \& Pauldrach (1991). 

The non detection of CO band heads in AG Carinae  could also be explained by a 
similar
mechanism. AG Carinae  was observed three times:
in March 1984, June 1984 and January 1985, always obtaining upper limits to the
flux determination of the three CO bands.  When comparing with the optical 
light curve
provided by Leitherer et al. (1994, their Figure 1), we find that AG Car  was 
transitioning
from maximum to minimum in 1982 - 1985, reaching a stable minimum at the 
beginning
of  1985. It is unfortunate that no subsequent CO observations were taken.

\section{Conclusions}

We have presented  the  first detection of $^{12}$CO J=2$\rightarrow$1 and
$^{12}$CO  J=1$\rightarrow$0   emission from the LBV AG Carinae. The emission 
is 
spatially unresolved.  We believe it to be spatially localised, rather than
extended,  and most likely associated with the immediate circumstellar region 
of AG
Carinae, i.e. inside the optical nebula. Both the detected CO lines are  characterised by a pseudo-gaussian 
profile of
FWHM $\simeq$ 15 km s$^{-1}$, indicating  a slowly expanding region
of molecular gas in close proximity to the hot central star. We have explored 
two
possible scenarios to explain the observed profile: a circumstellar envelope, 
similar to carbon stars, or a circumstellar disk. We believe that the 
option
of the circumstellar disk is preferable to provide the shielding necessary for 
the CO molecules to survive.

What is the origin of the CO emission? The outflow velocity   ($\simeq$ 7 km 
s$^{-1}$)
is atypically low when compared to the wind velocities  of LBVs (V$_{\inf}$ $>
$ 100-200 km s$^{-1}$). As we already mentioned, the CO velocity is also much 
slower than the expansion velocity of the optical nebula ($\simeq$ 70 km 
s$^{-1}$), with whom it is unlikely associated.
In terms of absolute values,  slow winds are usually found  in red supergiants 
(RSG). The question whether AG Carinae spent  any time as  a RSG is still 
open and argued on the basis of independent observational facts. 

Smith et al. (1988) and Waters et al. (1998) argued that the abundances of the
nebula and the composition of the dust would be in agreement with the ejection of
the nebula as a RSG. Alternatively, Lamers et al.  (2001) have shown  that the
abundances, the velocity, the  dynamical age and the morphology of the nebula are
all consistent with the nebula being ejected  immediately after the main sequence
phase of a rapidly rotating star. In that case the star cannot have been a RSG in the evolutionary sense (i.e. a massive star with a very extended
convective  envelope), but it may have resembled a RSG during an outburst phase,
when the wind was  optically thick, the effective radius very large and the outflow
velocity very low.

We consider several possible origins for the CO emission: 
\begin{itemize}
\item In the scenario proposed by Smith et al. (1998) which includes a RSG 
phase,
or by Lamers et al. (2001) which includes a brief RSG-like phase,
the molecular outflow could have originated during the RSG phase and survived
protected by the disk likely formed when the star, evolving as an LBV to the 
blue
side of the HRD, developed its fast wind. This could easily explain the low 
expansion
velocity.
\item Alternatively, the CO may originate at the outer regions of a 
circumstellar disk,
similar to the case of the B[e] stars. In this case the low velocity is 
difficult to
explain. In normal B[e] stars the outflow velocity is of order of 100 km 
s$^{-1}$, even
at large distances as shown by the forbidden lines (Zickgraf 1986). This is 
much larger
than the value of 7 km s$^{-1}$ that we observe.
\item Or, the CO emission may originate at the interaction-zone at the edge of 
the IS
bubble that was blown by the star. The maximum size of the emitting region 
($<$ 1 arcmin)
corresponds to a radius of about 1 pc. The amount of mass lost by the star in 
its main sequence
phase is about 10 M$_{\odot}$. The observed amount of molecular gas is 
$\simeq$ 3 M$_{\odot}$,
which is only about 1/4 of the total mass of the shocked gas at the edge of 
the bubble.
The main problem with this interpretation is  the small radius of the 
unresolved CO-emitting region (diameter $<$ 30$''$ or 0.9 pc) compared
with the predicted much larger  size (several tens of parsecs) of the bubble blown
by a spherical wind from a massive star with a velocity of 1000 km s$^{-1}$ during
4 $\times$ 10$^6$ years
(see e.g. Lamers \& Cassinelli 1999, p. 369). However,
if the wind were aspherical, e.g. concentrated in a disk due to the rapid 
rotation,
the expansion speed of the interaction region in the equatorial plane would be 
much
smaller because the bubble can more easily expand in the other direction. This 
might
result in a high density and very low velocity {\it waist} of the bipolar 
interstellar bubble, from
which the observed CO emission might originate.
\end{itemize}

We estimate a lower limit to the mass of the molecular region of 2.8 
M$_{\odot}$.
This is smaller, but still comparable with the mass of ionized gas present in 
the circumstellar
environment (4.2 M$_{\odot}$). The  implication  is that the molecular gas 
fraction can contribute significantly to the overall  mass lost from the 
central
star in its post main sequence evolution, and therefore one 
should be careful when assuming that the ionized gas mass well approximates, 
at least
in the case of AG Carinae, the total mass lost.

\acknowledgments

APM  would like to acknowledge financial support from NASA ADP grants
NAG5-2999 and NAG5-6854 and a grant from NASA administered by the AAS.
APM and HJGLML would also like to thank the STScI's Visitor Fund  for the funding 
and the hospitality.

\newpage

{\bf FIGURES}

{\bf Figure 1:}A 30$^{\prime}$ x 30$^{\prime}$ optical image  of the region around AG Carinae
(in the center) shows the 
position of the  $^{12}$CO J=1$\rightarrow$0 observations. The size of the
symbols is
matched to the size of the beam. North is up and East is to the left.
\\

{\bf Figure 2:} H$\alpha$ image of the ejected nebula surrounding AG Carinae
 on which we overlaid the pointings
for the $^{12}$CO J=2$\rightarrow$1 map, which are spaced by 12$''$. The circle indicates the beam
width for these observations (23$''$ in diameter).  In this image,
 North is up, East is to the left.
\\

{\bf Figure 3:} The $^{12}$CO J=1$\rightarrow$0 and $^{12}$CO J=2$\rightarrow$1
profiles are shown for the position centered on the star AG Carinae. A
multi-gaussian fit has been performed for each observed line.
\\

{\bf Figure 4:} Map for the $^{12}$CO J=1$\rightarrow$0 observations of the region
 around AG Carinae.
The numbers in abscissa/ordinate provide the pointing location in arcminutes from
the star (N, S, E or W). 
\\

{\bf Figure 5:} Radial velocity map of the  two narrow components in the $^{12}$CO
J=1$\rightarrow$0 emission as a function of position with respect to the star. 
The full dots and the empty dots identify the two components.
\\

{\bf Figure 6:}The mean interstellar, broad component (dashed line) is superposed 
to
the $^{12}$CO J=1$\rightarrow$0 profile observed on the star. The `on star' 
broad
$^{12}$CO J=1$\rightarrow$0 emission is  a factor of 2 higher than its 
interstellar counterpart, and it is slightly broader.
\\

{\bf Figure 7:} Map for the $^{12}$CO J=2$\rightarrow$1 observations of the region
around AG Carinae. The numbers in abscissa/ordinate provide the pointing
location in arcseconds from the star (N, S, E or W). 
\\

{\bf Figure 8:} Gaussian multifit to the various components present in the 
$^{12}$CO J=2$\rightarrow$1
profile on the star (upper panel). The various components are identified and 
tracked
to a distance of 50$''$ from the star. They move in parallel sheets, and they 
show a limited
spatial extent. Most likely, they belong to the underlying molecular gas 
region.

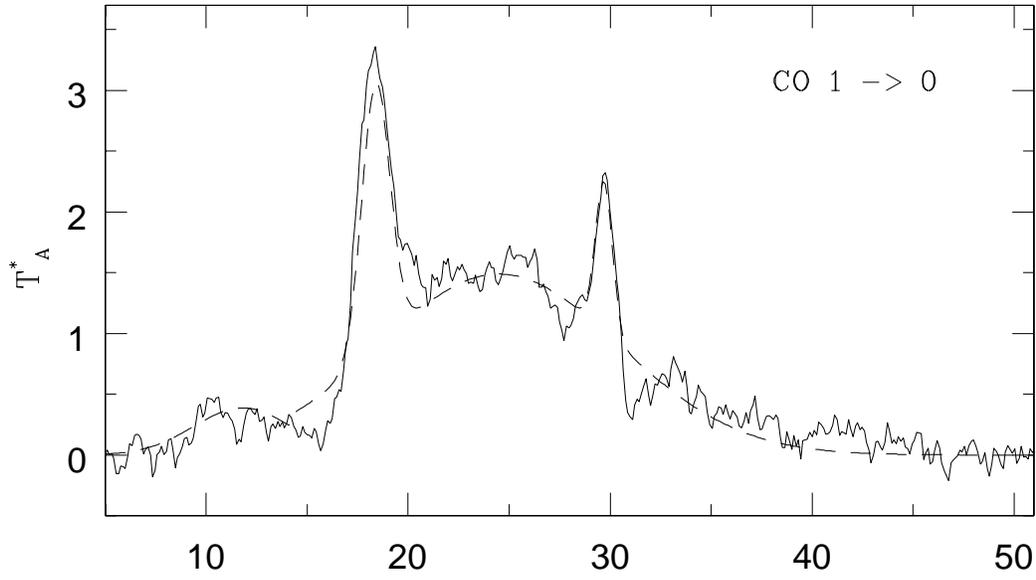
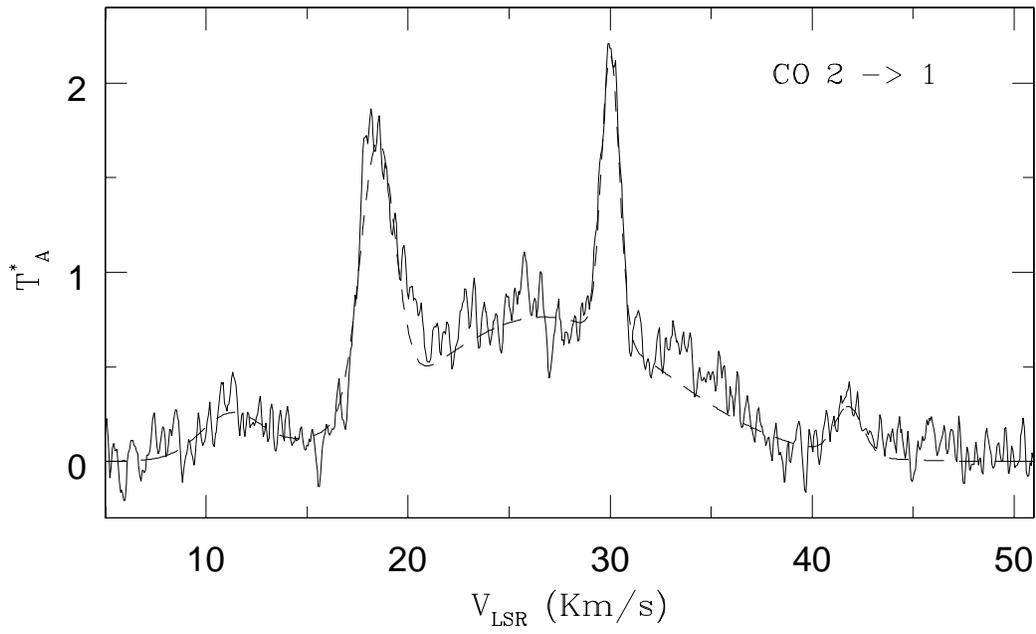
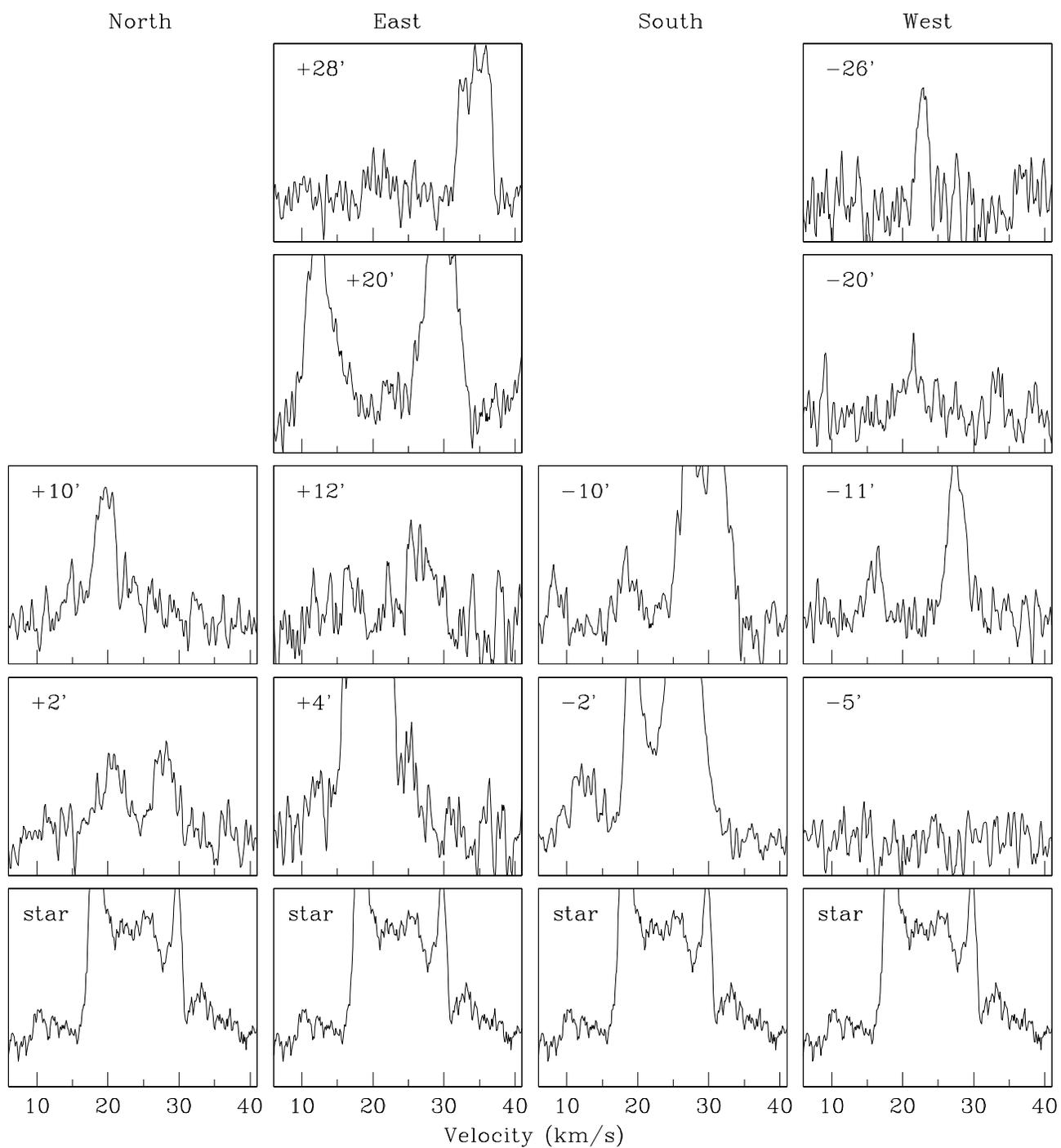
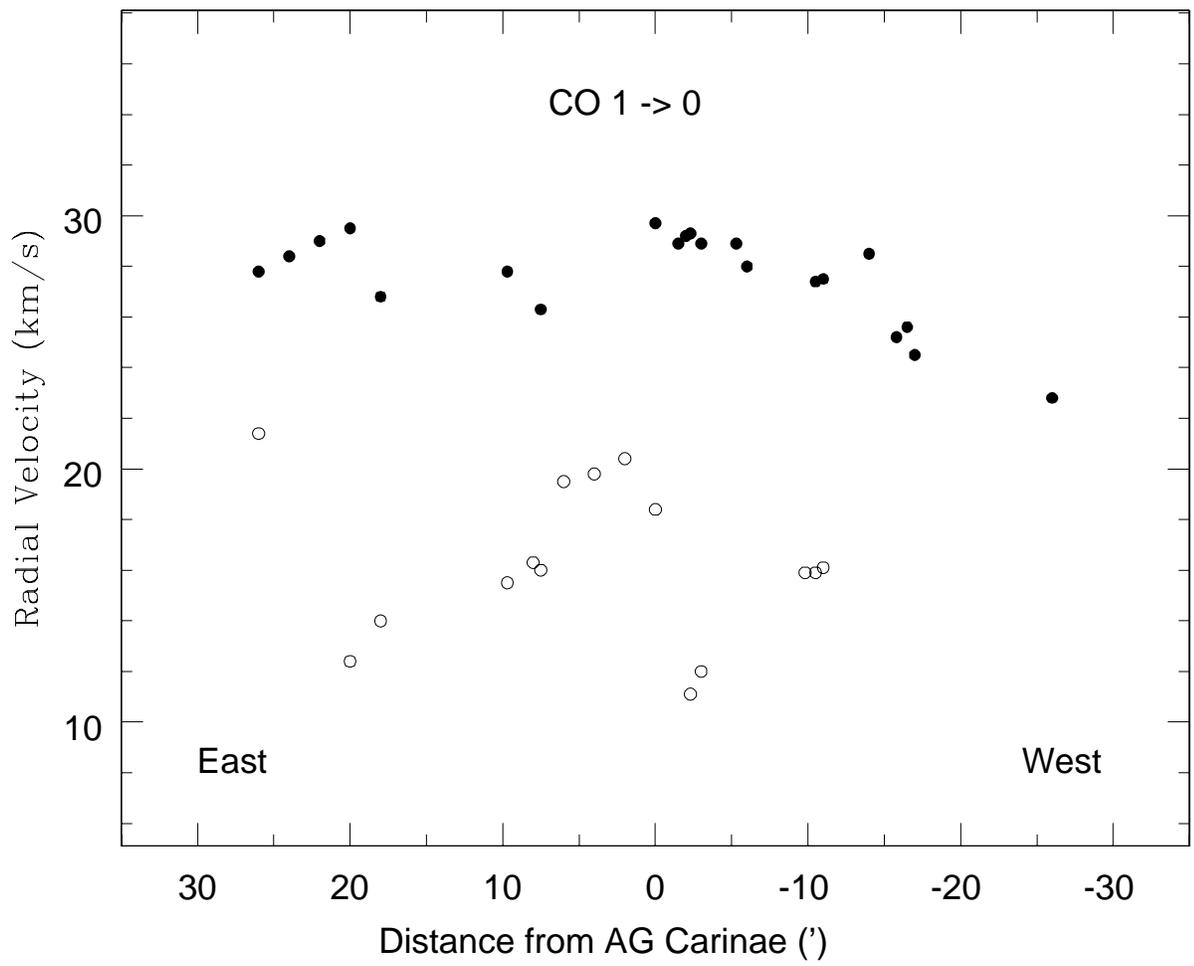
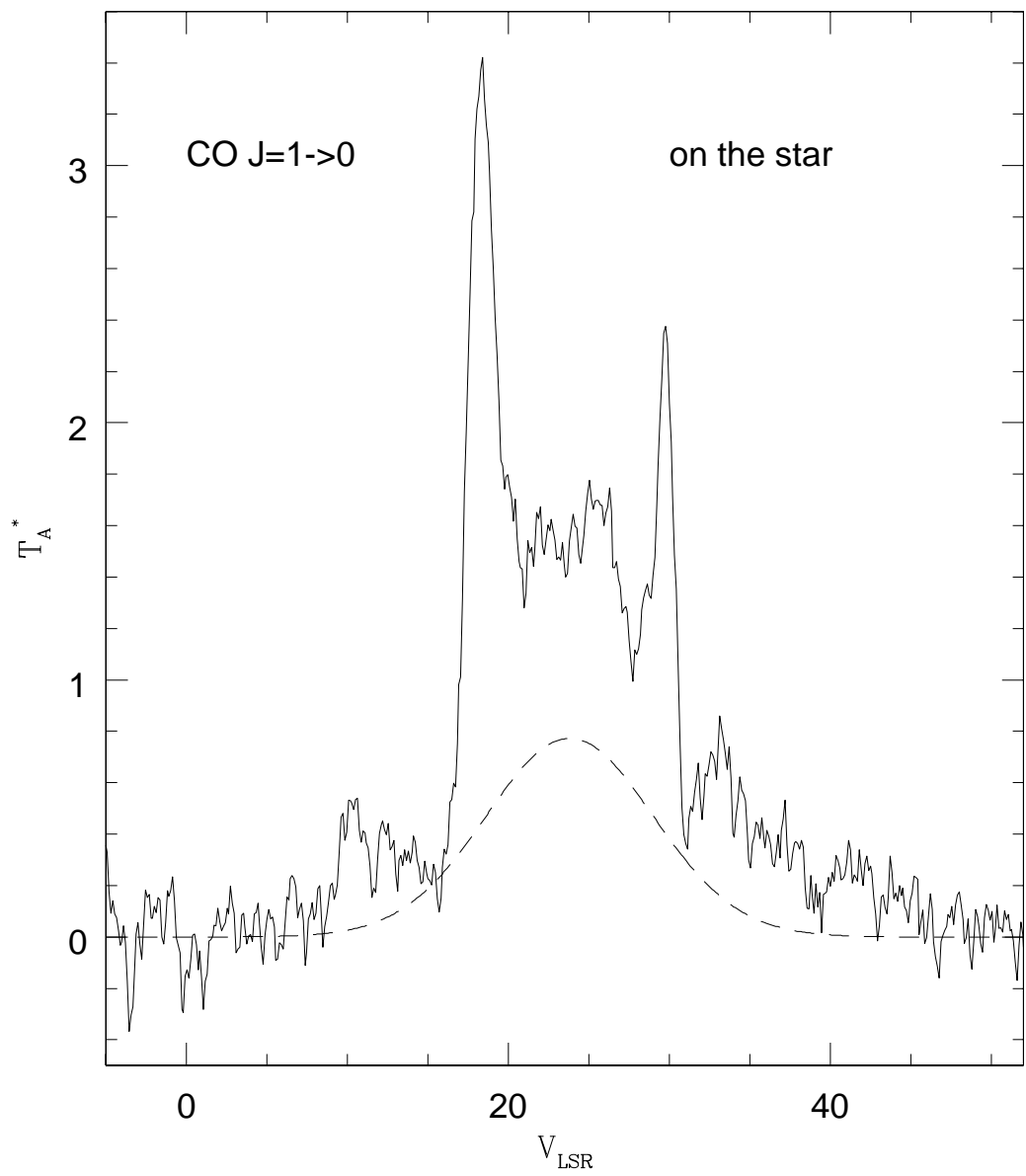
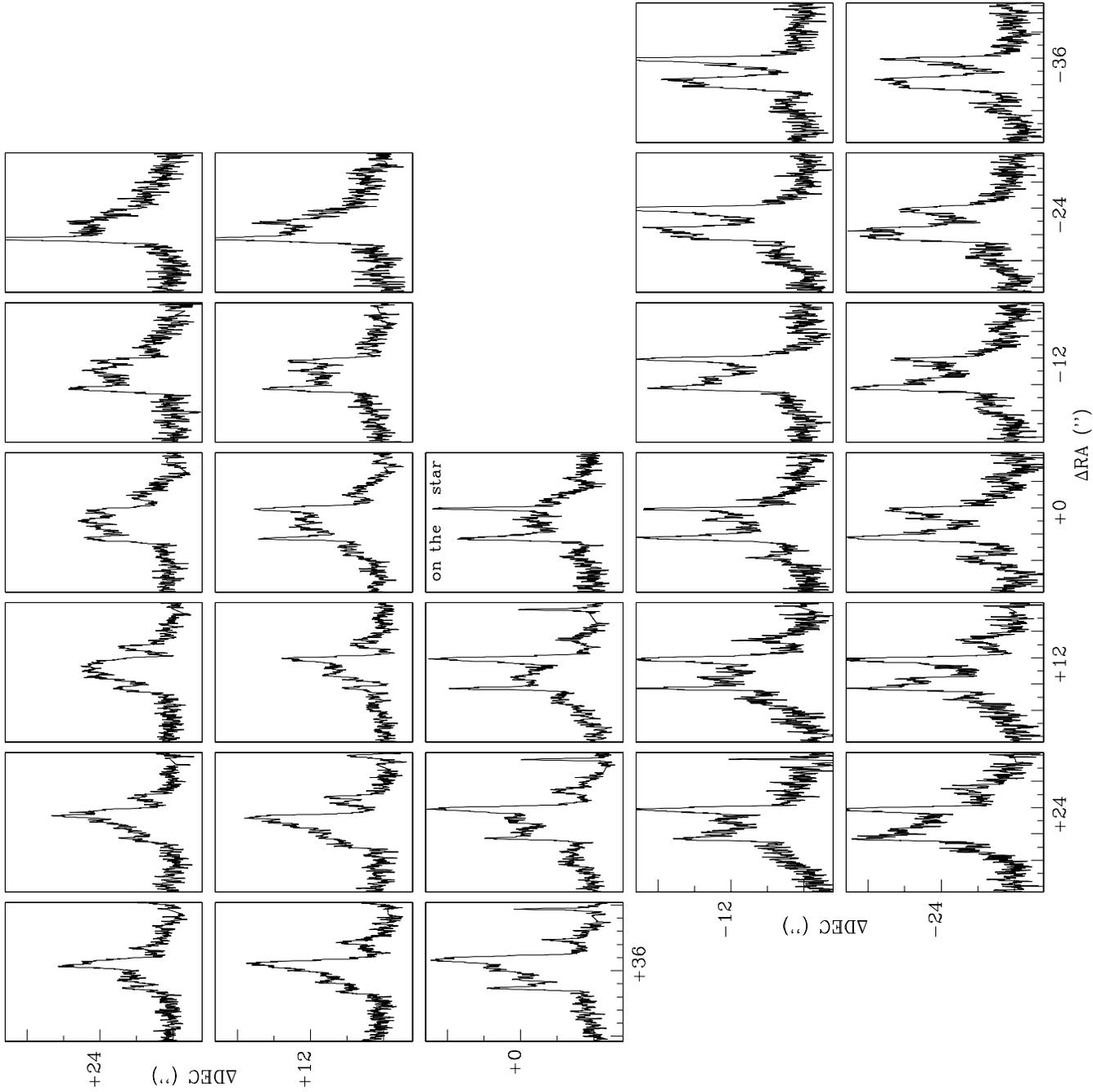
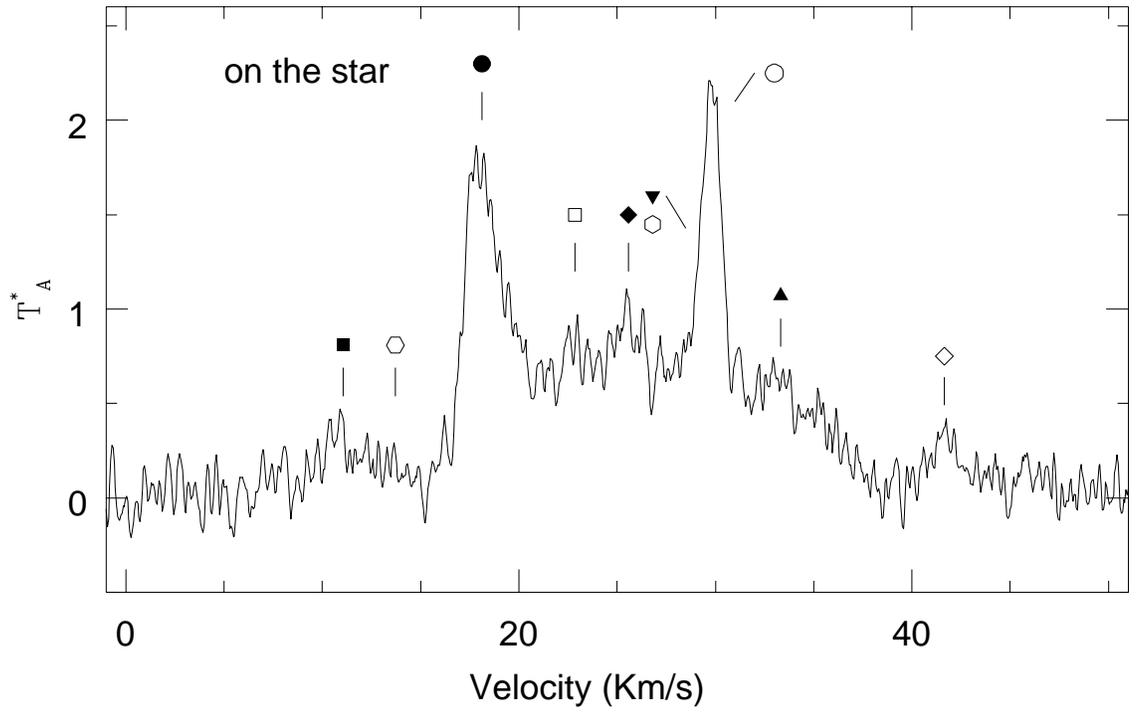
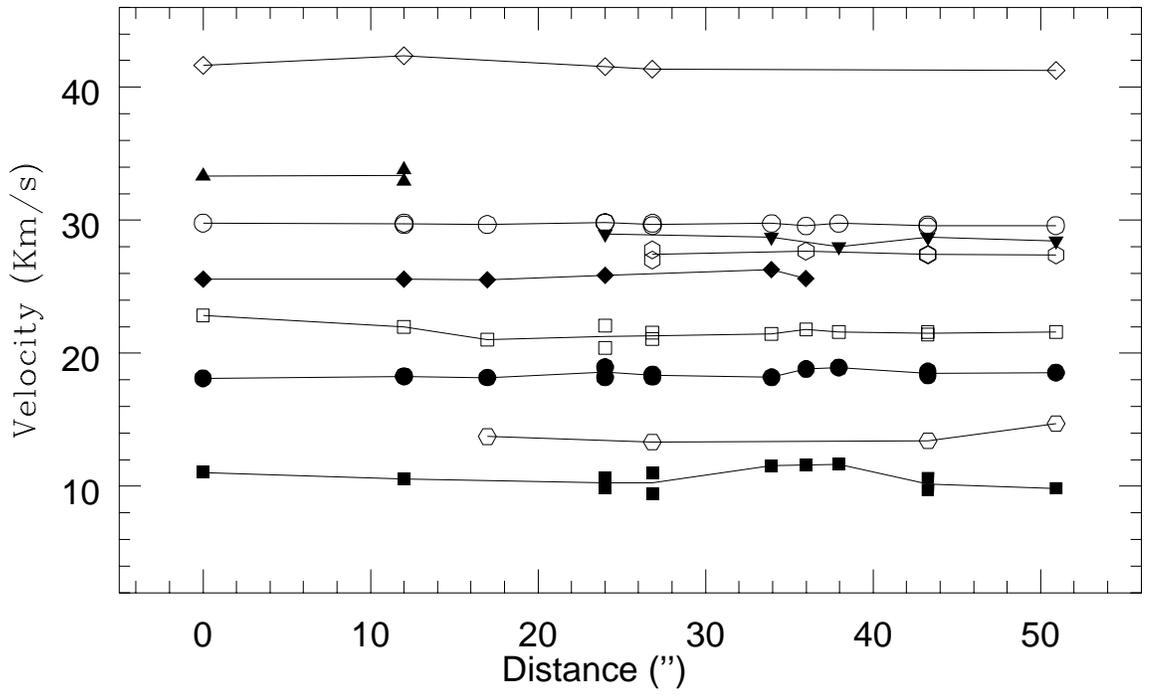
\end{document}